\documentclass[prl,twocolumn,aps,floatfix,showpacs,tightenlines,superscriptaddress,amsmath,amssymb,showkeys,10pt,nofootinbib,longbibliography]{revtex4-1}

\usepackage{graphicx}
\usepackage{amsmath}
\usepackage{amsfonts}
\usepackage{epsfig}
\usepackage{hhline}
\usepackage{soul}
\usepackage{tabularx,pifont,multirow}
\usepackage{txfonts, dsfont}
\usepackage{epsfig,libertine}
\usepackage[libertine]{newtxmath}
\usepackage{enumitem}
\usepackage{colordvi}
\usepackage{dsfont}
\usepackage{gensymb}

\usepackage[pdftex,
            pdfauthor={Kyrylo Simonov},
            pdftitle={Observability of spontaneous collapse in flavor oscillations and its relation to the CP and CPT symmetries}]{hyperref}
\hypersetup{colorlinks=true,urlcolor=red,citecolor=blue,filecolor=magenta,bookmarks=true}

\begin{document}

\title{Observability of spontaneous collapse in flavor oscillations and its relation to the $\mathcal{CP}$ and $\mathcal{CPT}$ symmetries}

\author{Kyrylo Simonov}
\email{kyrylo.simonov@univie.ac.at}
\affiliation{s7 rail technology GmbH, Lastenstra\ss e 36, 4020 Linz, Austria}
\affiliation{Fakult\"{a}t f\"{u}r Mathematik, Universit\"{a}t Wien, Oskar-Morgenstern-Platz 1, 1090 Vienna, Austria}

\bigskip 

\begin{abstract}
Spontaneous collapse models aim at solving the measurement problem of quantum mechanics by introducing collapse of wave function as an ontologically objective mechanism that suppresses macroscopic superpositions. In particular, the strength of collapse depends on the mass of the system. Flavor oscillating systems such as neutral mesons feature superpositions of states of different masses and, hence, could be used to test the validity of spontaneous collapse models. Recently, it has been shown that the mass-proportional CSL model causes exponential damping of the neutral meson oscillations which, however, is not strong enough to be observed in the present accelerator facilities. In this Letter, we study how the violation of the $\mathcal{CP}$ symmetry in mixing changes the spontaneous collapse effect on flavor oscillations and its observability.
\end{abstract}

\maketitle

\emph{Introduction.}---
Breakdown of the superposition principle of quantum mechanics at the macroscopic scale and subsequent emergence of classicality have been a subject of debates since the birth of quantum theory. Spontaneous collapse models attempt to explain it by adding an objective mechanism that stochastically reduces the wave function of a system~\cite{Bassi2003, Bassi2012}. For microscopic objects, the reduction mechanism's effect is negligible, so that the Schr\"odinger unitary evolution and the superposition principle remain valid. However, moving towards macroscopic scale, it is amplified and effectively suppresses superpositions of macroscopic states. In this way spontaneous collapse establishes a border between macroscopic and microscopic. 

The validity of spontaneous collapse models is a subject of intense experimental verification~\cite{CarlessoDonadi, Carlesso2022}. Fundamental testability of spontaneous collapse allows for a number of experimental proposals such as X-rays~\cite{Curceanu2015, Piscicchia2015, Curceanu2016, Piscicchia2017, Curceanu2019}, spontaneous radiation emission from charged particles~\cite{BassiDuerr, BassiDonadi, Adler2013, Donadi2021}, cold-atom experiments~\cite{Bilardello2016}, gravitational waves~\cite{Carlesso2016}, levitated nanoparticles~\cite{Vinante2019}, matter-wave interferometry~\cite{Nimmrichter2011, Toros2017, Toros2018, Schrinski2020}, and optomechanical setups~\cite{Carlesso2018, Nobakht2018, Carlesso2019}. Another example is given by mixed particle systems~\cite{Donadi2012, Bahrami2013, Donadi2013, SimonovLetter2016, SimonovPaper2016, Simonov2018, Simonov2020} that reveal phenomena of mixing and flavor oscillations. Because of these phenomena the states of a mixed particle which are relevant in experiments are given by superpositions of states with distinct masses and, therefore, are interesting for quantum-mechanical tests~\cite{Capolupo2019, Simonov2019, Naikoo2020, Buoninfante2020, Blasone2021}. A particular interest is drawn by neutral K-meson systems that violate the $\mathcal{CP}$ discrete symmetry and offer a rich playground for tests of spontaneous collapse models and foundations of quantum mechanics in general~\cite{Genovese2004, Bramon2006, Bernabeu2006, Hiesmayr2008, Yabsley2008, Blasone2008, AmelinoCamelia2010, DiDomenico2012, Hiesmayr2012, Bernabeu2012, Durt2019, Bernabeu2019, Blasone2021}.

\emph{Collapse models.}---
In general, the collapse dynamics is described by the following non-linear stochastic modification of Schr\"odinger equation,
\begin{eqnarray}
\nonumber d|\psi_t\rangle &=& \Bigr[
-i\hat{H}dt + \sqrt{\lambda}\sum_{i=1}^N (\hat{A}_i - \langle \hat{A}_i\rangle_t) dW_{i,t} \\
&-& \frac{\lambda}{2} \sum_{i=1}^N (\hat{A}_i - \langle \hat{A}_i\rangle_t)^2 dt \Bigr] |\psi_t\rangle, \label{CollapseEq}
\end{eqnarray}
where $\langle \hat{A}_i\rangle_t =  \langle\psi_t|\hat{A}_i |\psi_t\rangle$. Here $\hat{H}$ is the standard Hamiltonian of the quantum system, $\hat{A}_i$ are a set of $N$ operators introducing the collapse in a certain basis choice (position basis in most cases). $W_{i,t}$ represent a set of independent standard Wiener processes and $\lambda$ is a constant introduced by the model which quantifies the strength of the collapse. Notice that we use natural units, i.e., $\hbar = c = 1$. For the sake of simplicity, we focus on GRW-type models~\cite{GRW}, where collapse is triggered by interaction with an external classical stochastic field, and choose the mass-proportional CSL model~\cite{PearleCSL, GhirardiCSL, Ghirardi1995} which is intensively studied in flavor-oscillating~\cite{Donadi2012, Donadi2013, Bahrami2013} and radiation emitting systems~\cite{BassiDuerr, BassiDonadi, Adler2013, Curceanu2015, Piscicchia2015, Curceanu2016, Piscicchia2017, Curceanu2019}. However, a similar analysis can be performed straightforwardly for the QMUPL model~\cite{QMUPL} as well.

The mass-proportional CSL model is defined by the collapse operator of the form
\begin{equation}
    \hat{A}_{\mathbf{x}} = \int d\mathbf{y} g(\mathbf{y} - \mathbf{x}) \sum_i \frac{m_i}{m_0} \hat{\psi}_i^\dagger(\mathbf{y}) \hat{\psi}_i(\mathbf{y}),
\end{equation}
where $\hat{\psi}_i^\dagger(\mathbf{y})$ and $\hat{\psi}_i(\mathbf{y})$ are the creation and annihilation operators of a particle of type $i$ and mass $m_i$ in a point $\mathbf{y}$, respectively, whereas $m_0 \approx 9.4 \cdot 10^2 \; \mathrm{MeV/c^2}$ is the reference mass equal to the mass of a nucleon. For the strength $\lambda$ of collapse in the CSL model two different values were proposed,
\begin{equation}\label{GRW_Lambda}
    \lambda = 10^{-36} \; \mathrm{m}^3 \mathrm{s}^{-1}   
\end{equation}
by Ghirardi, Pearle, and Rimini~\cite{GhirardiCSL}, and
\begin{equation}
    \lambda = 10^{-28} \; \mathrm{m}^3 \mathrm{s}^{-1}   
\end{equation}
by Adler~\cite{Adler2007}. In what follows, due to the latest results of non-interferometric experiments probing spontaneous collapse~\cite{Carlesso2022}, we stick to the value (\ref{GRW_Lambda}). The function $g(\mathbf{y} - \mathbf{x})$ is the Gaussian smearing function
\begin{equation}
    g(\mathbf{y} - \mathbf{x}) = \frac{1}{(\sqrt{2\pi}r_C)^3} e^{-\frac{\mathbf{x}^2}{2r_C^2}},
\end{equation}
that introduces the second constant
\begin{equation} \label{rC}
r_C = 10^{-7} \; \mathrm{m}    
\end{equation}
of the CSL model setting the spatial scale of spontaneous collapse. Notice that the index $\mathbf{x}$ of the collapse operator is continuous and corresponds to a point in the physical space, hence, the sums in~(\ref{CollapseEq}) have to be replaced by integrals on $\mathbf{x}$.

\emph{Neutral meson dynamics and discrete symmetries.}---
Mixing and oscillations of neutral mesons are usually treated via non-relativistic quantum mechanics: therein, a neutral meson system is described by a two-dimensional Hilbert space $\mathbf{H}_M$ spanned by its physical particle/antiparticle states $|M^0\rangle$ and $|\bar{M}^0\rangle$ that are labeled by a flavor quantum number (for example, strangeness in the case of neutral $K$-mesons). The standard approach to dynamics is based on the Wigner--Weisskopf approximation which incorporates the decay into neutral meson dynamics by introducing the effective non-Hermitian Hamiltonian (written here in the $|M^0/\bar{M}^0\rangle$-basis)
\begin{eqnarray}
    \nonumber \hat{H}_{WWA} &=& \hat{M} - \frac{i}{2}\hat{\Gamma} \\
    &=& \begin{pmatrix} M_{11} - \frac{i}{2}\Gamma_{11} & M_{12} - \frac{i}{2}\Gamma_{12} \\ M_{12}^* - \frac{i}{2}\Gamma_{12}^* & M_{22} - \frac{i}{2}\Gamma_{22} \end{pmatrix} \equiv \begin{pmatrix} H_{11} & H_{12} \\ H_{21} & H_{22} \end{pmatrix}, \label{nonHermHam}
\end{eqnarray}
where $\hat{M} = \hat{M}^\dagger$ represents the mass operator which covers the unitary part of the dynamics, and $\hat{\Gamma} = \hat{\Gamma}^\dagger$ describes the decay. Conservation of discrete symmetries implies certain constraints on the elements of the Hamiltonian~(\ref{nonHermHam}). In particular, $\mathcal{CPT}$ symmetry requires $H_{11} = H_{22}$, while $\mathcal{CP}$ symmetry adds $|H_{12}| = |H_{21}|$ to this condition~\cite{Capolupo2011}.

Despite its wide usage in neutral meson phenomenology, the usage of Schr\"odinger equation with the Hamiltonian~(\ref{nonHermHam}) has to be justified. First of all, since (\ref{nonHermHam}) has a non-Hermitian part covering the decay, the resulting temporal part of the neutral meson evolution is not normalized\footnote{Interestingly, its possible renormalization by the decay width would cure this problem, however, it is ruled out by the $\mathcal{CP}$ violation in neutral K-mesons~\cite{Durt2019} highlighting the crucial role of fundamental discrete symmetries in quantum mechanics.}, and the corresponding probability of particle detection is not conserved~\cite{Bertlmann2006}. This leads to a natural question whether the evolution under the non-Hermitian Hamiltonian (\ref{nonHermHam}) has indeed a consistent probabilistic description. It can be demonstrated that the decay property can be consistently incorporated by treating decaying particles as an open system via the Gorini--Kossakowski--Lindblad--Sudarshan (GKLS) master equation in a larger Hilbert space $\mathbf{H}_M \oplus \mathbf{H}_D$, where $\mathbf{H}_D$ describes the decay products~\cite{Bertlmann2006, Bernabeu2012},
\begin{eqnarray}\label{EnlargedME}
    \frac{d\hat{\varrho}}{dt} &=& -i[\hat{\mathcal{M}}, \hat{\varrho}] - \frac{1}{2} \Bigl( \hat{\mathcal{B}}^\dagger\hat{\mathcal{B}} \hat{\varrho} + \hat{\varrho} \hat{\mathcal{B}}^\dagger\hat{\mathcal{B}} - 2 \hat{\mathcal{B}}\hat{\varrho}\hat{\mathcal{B}}^\dagger\Bigr),
\end{eqnarray}
where $\hat{\varrho} = \begin{pmatrix} \hat{\rho}_M & \hat{\rho}_{MD} \\ \hat{\rho}^\dagger_{MD} & \hat{\rho}_D \end{pmatrix}$ is a state in $\mathbf{H}$, while $\hat{\rho}_M = \hat{\rho}_M^\dagger$ and $\hat{\rho}_D = \hat{\rho}_D^\dagger$ are its components in $\mathbf{H}_M$ and $\mathbf{H}_D$, respectively. The unitary part of neutral meson dynamics is governed by the mass operator $\hat{\mathcal{M}} = \begin{pmatrix} \hat{M} & 0 \\ 0 & 0 \end{pmatrix}$ acting only in $\mathbf{H}_M$, whereas $\hat{\mathcal{B}} = \begin{pmatrix} 0 & 0 \\ \hat{B} & 0 \end{pmatrix}$ with $\hat{B} = \sum_{ij} \gamma_{ij} |f_i\rangle \langle M_j|$, where $\{ |f_i\rangle \}$ and $\{ |M_j\rangle \}$ are certain orthonormal bases in $\mathbf{H}_D$ and $\mathbf{H}_M$, respectively, are Lindblad operators describing transitions from $\mathbf{H}_M$ to $\mathbf{H}_D$, hence, triggering the decay. Normalizing the components of $\hat{B}$ as $\sum_i \gamma^*_{ij} \gamma_{ij'} = \delta_{jj'} \Gamma_j$, one obtains the following decomposition of (\ref{EnlargedME}) with respect to $\mathbf{H}_M$ and $\mathbf{H}_D$,
\begin{eqnarray}\label{Hamiltonian}
    \frac{d\hat{\rho}_M(t)}{dt} &=& -i[\hat{M}, \hat{\rho}_M(t)] - \frac{1}{2}\{\hat{B}^\dagger \hat{B}, \hat{\rho}_M(t)\}, \\
    \frac{d\hat{\rho}_D(t)}{dt} &=& \hat{B} \hat{\rho}_M(t) \hat{B}^\dagger, \\
    \frac{d\hat{\rho}_{MD}(t)}{dt} &=& -i\hat{M} \hat{\rho}_{MD}(t) - \frac{1}{2} \hat{B}^\dagger \hat{B}\hat{\rho}_{MD}(t),
\end{eqnarray}
with $\operatorname{Tr}[\hat{\rho}_M(t)] + \operatorname{Tr}[\hat{\rho}_D(t)] = 1$ due to the total GKLS equation~(\ref{EnlargedME}). First of all, the evolution of the off-diagonal element $\hat{\rho}_{MD}(t)$ is independent of $\hat{\rho}_{M}(t)$ and $\hat{\rho}_{D}(t)$ and can be ignored: in accelerator facilities, neutral mesons are produced in a certain state belonging to $\mathbf{H}_M$ meaning that $\hat{\rho}_D(t=0) = \hat{\rho}_{MD}(t=0) = 0$ and $\hat{\rho}_{MD}(t)$ remains zero. More importantly, the evolution of the decay counterpart $\hat{\rho}_D(t)$ is completely defined by the evolution of the flavor counterpart $\hat{\rho}_M(t)$, which, in turn, is equivalent to Schr\"odinger evolution under the effective Hamiltonian~(\ref{nonHermHam}),
\begin{equation}\label{EqWWA}
    \frac{d\hat{\rho}_M(t)}{dt} = -i[\hat{M}, \hat{\rho}_M(t)] - \frac{1}{2}\{\hat{\Gamma}, \hat{\rho}_M(t)\},
\end{equation}
with the decay operator $\hat{\Gamma} = \hat{B}^\dagger \hat{B} = \sum_i \Gamma_i |M_i\rangle \langle M_i|$. Crucially, it is proven that both decay and flavor counterparts are positive, and, moreover the evolution of the latter is \textbf{completely positive}~\cite{Bertlmann2006}. This means that $\hat{\rho}_M(t)$ guarantees correct probabilistic description of transitions between states in $\mathbf{H}_M$, and we can stick to the master equation~(\ref{EqWWA}) whose solution defines the total evolution in $\mathbf{H}$.

\emph{Violation of the $\mathcal{CP}$ symmetry.}---
When the $\mathcal{CP}$ symmetry is conserved (i.e., $|H_{12}| = |H_{21}|$, the eigenstates $|M_i\rangle$ of (\ref{nonHermHam}) have well-defined (distinct) definite masses $m_i$ and known decay widths $\Gamma_i$, so that the corresponding eigenvalues read $\lambda_i = m_i - \frac{i}{2}\Gamma_i$. For neutral mesons, there are two eigenstates labelled by $i = L, H$ corresponding to ``light'' and ``heavy'', that have certain mass difference $\Delta m = m_H - m_L$ and decay width difference\footnote{Traditionally, in meson phenomenology, the decay width difference is defined as $\Delta\Gamma = \Gamma_H - \Gamma_L$. However, in what follows, we focus only on neutral K-mesons, which are known to have $\Delta\Gamma < 0$. Therefore, for the sake of simplicity, we invert the definition of $\Delta\Gamma$ in order to operate with a positive value.} $\Delta\Gamma = \Gamma_L - \Gamma_H$ (for neutral K-mesons, the flavor states $|M^0/\bar{M}^0\rangle$ are denoted as $|K^0/\bar{K}^0\rangle$, while the mass eigenstates $|M_{L/H}\rangle$ correspond to $|K_{S/L}\rangle$ meaning ``short-lived'' and ``long-lived'', respectively). However, when a violation of the $\mathcal{CP}$ symmetry is taken into account, so that $|H_{12}| \neq |H_{21}|$, $\hat{H}_{WWA}$ is not normal since the decay and mass operators do not commute anymore, $[\hat{M}, \hat{\Gamma}] \neq 0$, preventing existence of states with both definite masses $m_{H/L}$ and decay widths $\Gamma_{H/L}$ and leading to non-orthogonality of $|M_{H/L}\rangle$,
\begin{equation}\label{NonOrth}
    \langle M_H | M_L \rangle \equiv \delta \neq 0.
\end{equation}
Neutral K-mesons are known to reveal a tiny violation of the $\mathcal{CP}$ symmetry with $\delta \approx 3.23 \cdot 10^{-3}$~\cite{Christenson1964, PartDataBook2020}. In this case, the highlighted problem of non-normal Hamiltonian~(\ref{nonHermHam}) can be overcome by treating $\mathcal{CP}$ violation as a perturbation~\cite{Bernabeu2013}. In this approach, instead of using non-orthogonal eigenstates $|K_{L/S}\rangle$ of $\hat{H}_{WWA}$, one focuses on the orthonormal $\mathcal{CP}$-eigenbasis
    \begin{equation}\label{CPBasis}
        |K_{1,2}\rangle = \frac{1}{\sqrt{2}} \Bigl( |K^0\rangle \pm |\bar{K}^0\rangle \Bigr),
    \end{equation}
and treats in it both the mass $\hat{M}$ and decay $\hat{\Gamma}$ operators. The latter, taking into account the dominant decay of neutral mesons to two pions and assuming conservation of $\mathcal{CPT}$ symmetry and absence of the $\mathcal{CP}$ violation in decay, is given by
    \begin{equation}
        \hat{\Gamma} = \begin{pmatrix} 0 & 0 \\ 0 & \gamma \end{pmatrix},
    \end{equation}
where $\gamma$ is therefore the difference of the decay widths of $|K_2\rangle$ and $|K_1\rangle$. In turn, the eigenstates of the mass operator
\begin{equation}
    \hat{M} = \begin{pmatrix} M - \Re M_{12} & -i\Im M_{12} \\ i\Im M_{12} & M + \Re M_{12} \end{pmatrix}
\end{equation}
with definite masses $m'_i$ differ from the $\mathcal{CP}$-eigenstates~(\ref{CPBasis}) with definite decay widths as~\cite{Bernabeu2012}
\begin{eqnarray}\label{MassState1}
    |\mathcal{M}_1\rangle &=& i|K_1\rangle + \frac{|\varepsilon|}{\sin(\phi)} |K_2\rangle + O(|\varepsilon|^2), \\
    |\mathcal{M}_2\rangle &=& -i\frac{|\varepsilon|}{\sin(\phi)}|K_1\rangle + |K_2\rangle + O(|\varepsilon|^2), \label{MassState2}
\end{eqnarray}
where the parameters
\begin{eqnarray}
    \varepsilon &=& |\varepsilon|e^{i\phi} = \frac{\Im{M_{12}}}{\frac{\gamma}{2} + i\Delta m'}, \\
    \tan(\phi) &=& \frac{2\Delta m'}{\gamma},
\end{eqnarray}
characterize the $\mathcal{CP}$ violation~\cite{Ellis1996, Bernabeu2012}, with $\Delta m' = 2|M_{12}|$. Neutral K-mesons are known to violate the $\mathcal{CP}$ violation with $|\varepsilon| \approx 2.23 \cdot 10^{-3}$ and $\phi \approx 43.5 \degree$~\cite{PartDataBook2020}.

Having properly defined the mass $\hat{M}$ and decay $\hat{\Gamma}$ operators, one can solve equation (\ref{EqWWA}) for the flavor counterpart $\hat{\rho}_M = \begin{pmatrix} \rho_{11} & \rho_{12} \\ \rho_{12}^* & \rho_{22} \end{pmatrix}$ in the $\mathcal{CP}$-eigenbasis~(\ref{CPBasis}) perturbatively in powers of $|\varepsilon| \propto \Im M_{12}$,
\begin{equation}\label{PertDM}
    \hat{\rho}_M(t) = \hat{\rho}_M^{(0)}(t) + |\varepsilon|\hat{\rho}_M^{(1)}(t) + ...
\end{equation}
In what follows, we focus on the first-order contribution of the $\mathcal{CP}$ violation, i.e., linear in $|\varepsilon|$. Since the difference of masses $\Delta m'$ and decay widths $\gamma$ differ from the corresponding parameters $\Delta m$ and $\Delta\Gamma$ of Wigner-Weisskopf approximation by $|\varepsilon|^2$~\cite{Bernabeu2012}, they can be regarded as equal in what follows. In turn, the corresponding non-orthogonality (\ref{NonOrth}) of the eigenstates $|K_{L/S}\rangle$ of $\hat{H}_{WWA}$ is given by $\delta = 2\Re \varepsilon$.

\emph{Transition probabilities and their asymmetry.}---
Now we turn to the transition probabilities for K-mesons, which can be calculated using $\hat{\rho}_M(t)$, and their combinations that are measured at the accelerator facilities. In particular, the KLOE and KLOE-2 experiments by the DA$\Phi$NE $\Phi$-factory provides an experimental setup with low environmental effects which is widely used to probe fundamental symmetries~\cite{DiDomenico2010, AmelinoCamelia2010, Bernabeu2013, BalwierzPytko2013, Babusci2014, Anastasi2018, DiDomenico2020, Babusci2022}. Therein, neutral K-mesons are produced as correlated pairs in an antisymmetric Bell state $|I\rangle = \frac{1}{\sqrt{2}}(|K^0\bar{K}^0\rangle - |\bar{K}^0K^0\rangle)$. We are interested in combinations of probabilities of measuring a kaon $|K^0\rangle$ or antikaon $|\bar{K}^0\rangle$ at a certain time point after a pair of K-mesons has been produced in the state $|I\rangle$. More precisely, given a pair of K-mesons in the state $|I\rangle$ at $t=0$, we are interested in constructing two-particle transition probabilities~\cite{Bertlmann2003},
\begin{eqnarray}\label{2PartProb}
    P_{I \rightarrow F_1 F_2}(t_1, t_2) &=& \langle F_2| \rho_{M|F_1}(t_1, t_2) |F_2\rangle, \\
    \rho_{M|F_1}(t_1, t_2 = t_1) &=& \operatorname{Tr}_1[(|F_1\rangle\langle F_1| \otimes \mathds{1}) \rho_{MM}(t_1)], \label{OnePart}
\end{eqnarray}
of obtaining outcome $F_1$ after measurement of the flavor content (i.e., strangeness) of the first particle at $t_1 > 0$, and outcome $F_2$ after measurement of strangeness of the second particle at $t_2 > t_1 > 0$, with $\operatorname{Tr}_1$ being partial trace over the Hilbert space of the first particle. Therein, the state $\rho_{MM}(t) \in \mathcal{L}(\mathbf{H}_M \otimes \mathbf{H}_M)$ of both particles evolves in time until the strangeness of the first particle is measured at $t = t_1$. If it is found in a state $|F_1\rangle \in \{ |K^0\rangle, |\bar{K}^0\rangle\}\}$, the state of the second particle is given by the one-particle state~(\ref{OnePart}). It evolves further until its strangeness content is measured at a later moment of time $t = t_2 > t_1$. In what follows, we focus on probabilities of finding the particles in opposite flavor states, $P_{I \rightarrow K^0 \bar{K}^0}(t_1, t_2)$ and $P_{I \rightarrow \bar{K}^0 K^0}(t_1, t_2)$, and their asymmetry,
\begin{equation}\label{StandardCPT}
    \mathbb{A}(t_1,t_2) = \frac{P_{I \rightarrow K^0\bar{K}^0}(t_1, t_2) - P_{I \rightarrow \bar{K}^0K^0}(t_1, t_2)}{P_{I \rightarrow K^0\bar{K}^0}(t_1, t_2) + P_{I \rightarrow \bar{K}^0K^0}(t_1, t_2)}.
\end{equation}


Within standard quantum mechanics, a neutral K-meson experiences a genuinely Schr\"odinger evolution under the Hamiltonian $\hat{H}_{WWA}$, and, in turn, a pair of them evolves under two-particle Hamiltonian $\hat{H}' = \hat{H}_{WWA} \otimes \mathds{1} + \mathds{1} \otimes \hat{H}_{WWA}$, i.e., with no further interaction between the particles after they have been produced in the state $|I\rangle$. Assuming that the $\mathcal{CPT}$ symmetry is not conserved (hence, $H_{11} \neq H_{22}$), and its violation is quantified by the complex parameter $z = \frac{H_{11} - H_{22}}{\Delta m + \frac{i}{2}\Delta\Gamma}$, the transition probabilities of interest are given by
\begin{widetext}
\begin{eqnarray}
\nonumber P_{I\rightarrow K^0\bar{K}^0 / \bar{K}^0 K^0}(t_1, t_2) &=& e^{- \frac{\Gamma}{2}(t_1+t_2)}\Bigl[ \frac{1 + |z|^2}{2} \cosh\Bigl( \frac{\Delta\Gamma \Delta t}{2}\Bigr) \pm \Re z \sinh\Bigl( \frac{\Delta\Gamma \Delta t}{2}\Bigr) + \frac{1 - |z|^2}{2}\cos(\Delta m \Delta t) \pm \Im z \sin(\Delta m \Delta t) \Bigr],
\end{eqnarray}
\end{widetext}
where $\Delta t = t_2 - t_1$ is the difference of measurement times. Their asymmetry~(\ref{StandardCPT}) reads
\begin{equation}\label{AsymTwoQM}
    \mathbb{A}_{QM}(\Delta t) = \frac{2\Re{z} \sinh\Bigl(\frac{\Delta\Gamma \Delta t}{2}\Bigr) + 2\Im{z} \sin(\Delta m \Delta t)}{(1+|z|^2)\cosh\Bigl(\frac{\Delta\Gamma \Delta t}{2}\Bigr) + (1-|z|^2) \cos(\Delta m \Delta t)},
\end{equation}
which depends only on $\Delta t$. It can be spotted that, if $\Delta t \neq 0$, $\mathbb{A}_{QM}(\Delta t)$ is non-zero only if $z \neq 0$, i.e., the $\mathcal{CPT}$ symmetry is violated in mixing. Otherwise, its conservation leads to $\mathbb{A}_{QM}(\Delta t) = 0$ for any $t_1, t_2$. Notice that, due to the presence of entanglement between the particles, two-particle asymmetry term~(\ref{AsymTwoQM}) formally coincides with the asymmetry term
\begin{equation}
    \mathbb{A}^{single}_{QM}(\Delta t) = \frac{P_{K^0 \rightarrow K^0}(\Delta t) - P_{\bar{K}^0 \rightarrow \bar{K}^0}(\Delta t)}{P_{K^0 \rightarrow K^0}(\Delta t) + P_{\bar{K}^0 \rightarrow \bar{K}^0}(\Delta t)},
\end{equation}
which is constructed from single-particle probabilities
\begin{equation}
P_{A\rightarrow B}(\Delta t) = |\langle B| \hat{U}(\Delta t) |A\rangle |^2
\end{equation}
of measuring a K-meson (being produced in a state $|A\rangle$ at time $t_0$) in a state $|B\rangle$ at time $t_1 > t_0$ with $\Delta t = t_1 - t_0$, and it is a known witness of the $\mathcal{CPT}$ violation~\cite{Capolupo2011}.

\emph{Impact of the spontaneous collapse.}---
We aim to investigate the two-particle asymmetry term~(\ref{StandardCPT}) under the mass-proportional CSL model and compare it with its value~(\ref{AsymTwoQM}) given by standard quantum mechanics. In turn, the time evolution of the flavor component (i.e., belonging to $\mathbf{H}_M$) of a quantum state is changed by adding a Lindblad term to the master equation~(\ref{EqWWA}). Indeed, with included spontaneous collapse effect with respect to the state vector equation~(\ref{CollapseEq}), the completely positive evolution of a neutral meson system is governed by the GKLS equation
\begin{widetext}
\begin{eqnarray}\label{Lindblad}
    \frac{d\hat{\rho}_M(t)}{dt} &=& -i[\hat{M}, \hat{\rho}_M(t)] - \{\hat{\Gamma}, \hat{\rho}_M(t)\} - \frac{\lambda}{2} \! \int \!\! d\mathbf{x} \Bigl( \hat{A}_\mathbf{x}^2 \hat{\rho}_M(t) + \hat{\rho}_M(t) \hat{A}_\mathbf{x}^2 - 2 \hat{A}_\mathbf{x} \hat{\rho}_M(t) \hat{A}_\mathbf{x} \Bigr),
\end{eqnarray}
\end{widetext}
which describes the flavor counterpart of the total dynamics governed by the GKLS equation
\begin{eqnarray}\label{EnlargedME_CSL}
    \nonumber \frac{d\hat{\varrho}(t)}{dt} &=& -i[\hat{\mathcal{M}}, \hat{\varrho}(t)] - \frac{1}{2} \Bigl( \hat{\mathcal{B}}^\dagger\hat{\mathcal{B}} \hat{\varrho}(t) + \hat{\varrho}(t) \hat{\mathcal{B}}^\dagger\hat{\mathcal{B}} - 2 \hat{\mathcal{B}}\hat{\varrho}(t)\hat{\mathcal{B}}^\dagger\Bigr) \\
    &-& \frac{\lambda}{2} \int d\mathbf{x} \Bigl( \hat{\mathcal{A}}_\mathbf{x}^2 \hat{\varrho}(t) + \hat{\varrho}(t) \hat{\mathcal{A}}_\mathbf{x}^2 - 2 \hat{\mathcal{A}}_\mathbf{x} \hat{\varrho}(t) \hat{\mathcal{A}}_\mathbf{x} \Bigr),
\end{eqnarray}
with $\hat{\mathcal{A}}_\mathbf{x} = \begin{pmatrix} \hat{A}_\mathbf{x} & 0 \\ 0 & 0\end{pmatrix}$, i.e., the CSL collapse process distinguishes between states~(\ref{MassState1}) and (\ref{MassState2}) in $\mathbf{H}_M$ with different definite masses. We assume that the $\mathcal{CPT}$ symmetry is conserved (i.e., $H_{11} = H_{22}$), whereas the $\mathcal{CP}$ symmetry is broken (i.e., $|H_{12}| \neq |H_{21}|$). Taking into account that the strength of the $\mathcal{CP}$ violation $|\varepsilon|$ in K-meson system is tiny, we calculate the transition probabilities~(\ref{2PartProb}) under spontaneous collapse dynamics solving perturbatively (\ref{Lindblad}) via (\ref{PertDM}),
\begin{equation}
    P_{I \rightarrow F_1 F_2}(t_1, t_2) = P^{(0)}_{I \rightarrow F_1 F_2}(t_1, t_2) + |\varepsilon| P^{(1)}_{I \rightarrow F_1 F_2}(t_1, t_2) + ...
\end{equation}
We start with the zeroth perturbative order, i.e., with the $\mathcal{CP}$-symmetry being conserved. As expected, we recover the known two-particle probabilities reported in~\cite{Donadi2012},
\begin{widetext}
\begin{eqnarray}
\nonumber P^{(0)}_{I \rightarrow K^0\bar{K}^0}(t_1, t_2) &=& \frac{e^{-\gamma(t_1 + t_2)}}{4} \Bigl[ \cosh\Bigl(\frac{\gamma \Delta t}{2}\Bigr) + e^{-\frac{\Lambda}{2}(t_1 + t_2)} \cos(\Delta m' \Delta t) \Bigr], \\
P^{(0)}_{I \rightarrow \bar{K}^0 K^0}(t_1, t_2) &=& P^{(0)}_{I \rightarrow K^0\bar{K}^0}(t_1, t_2),
\end{eqnarray}
\end{widetext}
where $\Lambda = \frac{\lambda}{(2\sqrt{\pi}r_C)^3} \frac{(\Delta m')^2}{m_0^2} \approx 3.0 \cdot 10^{-46} \;\mathrm{s}^{-1}$ defines additional exponential damping of the flavor oscillations due to the mass-proportional CSL model. 
%
This makes the CSL effect on a neutral meson system with the conserved $\mathcal{CP}$ symmetry too weak with respect to the sensitivity of the present accelerator facilities including the DA$\Phi$NE collider~\cite{Bahrami2013, Donadi2012}. More interesting are the first-order corrections to the probabilities, where $\mathcal{CP}$ violation appears,
\begin{widetext}
\begin{eqnarray}
\nonumber P^{(1)}_{I \rightarrow K^0\bar{K}^0}(t_1, t_2) &=& \frac{|\varepsilon|\Lambda \frac{\gamma}{\Delta m'} \sqrt{\gamma^2 + 4(\Delta m')^2} \; e^{-\gamma(t_1 + t_2)}}{((\gamma - \Lambda)^2 + 4(\Delta m')^2)((\gamma + \Lambda)^2 + 4(\Delta m')^2)} \Biggl[ 4\gamma\Delta m' \Biggl(\sinh\Bigl(\frac{\gamma\Delta t}{2}\Bigr) - e^{-\frac{\Lambda}{2}t_1} \sinh\Bigl( \frac{\gamma}{2} t_2 \Bigr) \cos(\Delta m' t_1) \\
\nonumber &+& e^{-\frac{\Lambda}{2}t_2} \sinh\Bigl( \frac{\gamma}{2} t_1 \Bigr) \cos(\Delta m' t_2) \Biggr) +\Bigl(\gamma^2 - \Lambda^2 - 4(\Delta m')^2\Bigr) \Biggl( e^{-\frac{\Lambda}{2}(t_1 + t_2)} \sin(\Delta m' \Delta t) \\
\nonumber &+& e^{-\frac{\Lambda}{2}t_1} \cosh\Bigl( \frac{\gamma}{2} t_2 \Bigr) \sin(\Delta m' t_1) - e^{-\frac{\Lambda}{2}t_2} \cosh\Bigl( \frac{\gamma}{2} t_1 \Bigr) \sin(\Delta m' t_2)  \Biggr], \\
P^{(1)}_{I \rightarrow \bar{K}^0 K^0}(t_1, t_2) &=& - P^{(1)}_{I \rightarrow K^0\bar{K}^0}(t_1, t_2),
\end{eqnarray}
\end{widetext}
\begin{center}
\begin{figure}[t!]
\centering
\includegraphics[width=\columnwidth]{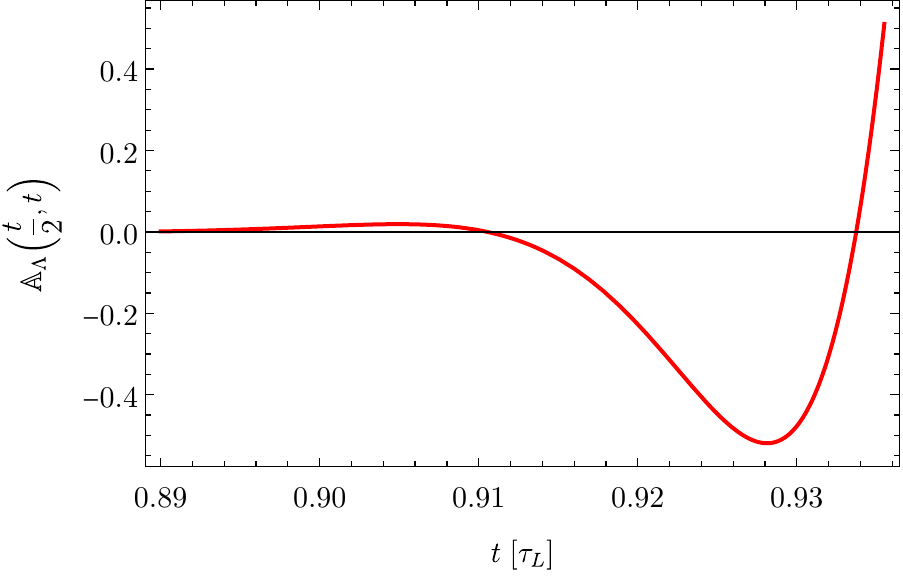}
\caption{The asymmetry term (\ref{CPT_CSL}) for the neutral kaon system, where the measurement times are assumed as $t_2 \equiv t$ and $t_1 = \frac{1}{2}t$ and given in units of the lifetime $\tau_L \approx 5.116 \cdot 10^{-8} \; \mathrm{s}$ of the long-lived state $|K_L\rangle$. The neutral kaon system is characterized by the difference of decay widths $\Delta\Gamma \approx 1.1149 \cdot 10^{10} \; \mathrm{s}^{-1}$ and mass difference $\Delta m \approx 0.5293 \cdot 10^{10} \; \hbar\mathrm{s}^{-1}$ and violates the $\mathcal{CP}$ symmetry with parameters $\delta \approx 3.23 \cdot 10^{-3}$ and $\phi \approx 43.5 \degree$~\cite{PartDataBook2020}. Consequently, the strength of the spontaneous collapse effect is characterized by $\Lambda = 3.0 \cdot 10^{-46} \mathrm{s}^{-1}$, where the GRW value (\ref{GRW_Lambda}) of the CSL constant $\lambda$ is considered.}
\label{CollapseRateFigure}
\end{figure}
\end{center}
The form of the first order contribution suggests that it turns zero when
\begin{itemize}
    \item $\Lambda = 0$, and no collapse is presented, or
    \item $\gamma = 0$, i.e., the $\mathcal{CP}$-eigenstates decay with the same rate, or
    \item $t_1 = t_2$, i.e., $\Delta t = 0$, so that both mesons are measured at the same point of time.
\end{itemize}
In contrast to the predictions~(\ref{AsymTwoQM}) of standard quantum mechanics, the transition probabilities under collapse dynamics do not have time translation invariance anymore, i.e., they depend not only on $\Delta t$ but the absolute times $t_{1,2}$ as well since spontaneous collapse dynamics corrupts the initial entanglement of the particles. This means that the effect of spontaneous collapse can be in principle distinguished from a violation of the $\mathcal{CPT}$ symmetry. Plugging in the obtained probabilities into~(\ref{StandardCPT}), we find
\begin{eqnarray}\label{CPT_CSL}
\nonumber \mathbb{A}_{\Lambda}(t_1, t_2) &=& |\varepsilon| \frac{P^{(1)}_{I \rightarrow K^0\bar{K}^0}(t_1, t_2)}{P^{(0)}_{I \rightarrow K^0\bar{K}^0}(t_1, t_2)}.
\end{eqnarray}
Taking into account that $\Delta m' \approx \Delta m$ and $\gamma \approx \Delta\Gamma$ as well as weakness of the spontaneous collapse rate $\Lambda$ and leaving only the terms linear in $\Lambda$, we obtain the final expression for the asymmetry term~(\ref{StandardCPT}) within spontaneous collapse dynamics,
\begin{widetext}
\begin{eqnarray}\label{CPT_CSL}
\nonumber \mathbb{A}_{\Lambda}(t_1, t_2) &\approx& \frac{2\delta\frac{\Lambda}{\Delta m}\sin(\phi)}{\cosh\Bigl(\frac{\Delta\Gamma \Delta t}{2}\Bigr) + \cos(\Delta m \Delta t)} \Biggl[ \sin(\phi) \Biggl( \sinh\Bigl(\frac{\Delta\Gamma\Delta t}{2}\Bigr) - \sinh\Bigl( \frac{\Delta\Gamma}{2} t_2 \Bigr) \cos(\Delta m t_1) + \sinh\Bigl( \frac{\Delta\Gamma}{2} t_1 \Bigr) \cos(\Delta m t_2) \Biggl)  \\
\nonumber &+& \cos(\phi) \Biggl( \sin(\Delta m \Delta t) + \cosh\Bigl( \frac{\Delta\Gamma}{2} t_2 \Bigr) \sin(\Delta m t_1) - \cosh\Bigl( \frac{\Delta\Gamma}{2} t_1 \Bigr) \sin(\Delta m t_2)\Biggr)\Biggr].
\end{eqnarray}
\end{widetext}

It can be seen that while increasing $\Delta\Gamma$ (for neutral K-mesons, $\Gamma_S \approx 600 \Gamma_L$ leading to $\Delta\Gamma \approx 1.1149 \cdot 10^{10} \; \mathrm{s}^{-1}$) the calculated asymmetry term (\ref{CPT_CSL}) becomes significant, first of all, when the flavors of both mesons are measured at times of the order of the long-lived meson lifetime $\tau_L = \Gamma_L^{-1} \approx 5.116 \cdot 10^{-8} \; \mathrm{s}$ (see Fig.~\ref{CollapseRateFigure}). This makes neutral K-meson system stand out from other neutral mesons whose differences of decay rates $\Delta\Gamma$ are negligible.

\emph{Conclusions.}---
In summary, the obtained result for the mass-proportional CSL model shows that even tiny violation of the $\mathcal{CP}$ symmetry can significantly change the dynamics of a decaying flavor oscillating system under spontaneous collapse. In particular, when the $\mathcal{CP}$ symmetry is broken, the CSL dynamics affects an asymmetry term witnessing $\mathcal{CPT}$ violation in standard quantum mechanics. This allows one to distinguish in principle between the effects of $\mathcal{CPT}$ violation and spontaneous collapse. Importantly, while increasing the difference $\Delta\Gamma$ between the decay widths, spontaneous collapse effect on a neutral meson system becomes stronger. Neutral K-mesons stand apart from other neutral meson systems because of significant difference of their decay widths: the difference of decay lengths reaches a magnitude of several centimeters. Hence, neutral kaons could provide a suitable setup to observe spontaneous collapse effect in the existing accelerator facilities. This suggests a further research of spontaneous collapse in a $\mathcal{CP}$-violating flavor oscillating system, in particular, how the effect of other types of collapse models (first of all, gravity-related collapse models~\cite{Karolyhazy1966, Diosi1984, Karolyhazy1986, Diosi1987, Penrose1996, Penrose1998, Donadi2021G} such as the Di\'osi-Penrose model) its dynamics differs from one of the CSL model.

\section*{Acknowledgements}
We thank Antonio Capolupo, Antonio Di Domenico, Salvatore Marco Giampaolo, and Beatrix Cornelia Hiesmayr for the numerous and fruitful discussions on physics of neutral mesons and spontaneous collapse models, and Aniello Quaranta and Luca Smaldone for the technical support. We acknowledge the support from the Austrian Science Fund (FWF-P26783, FWF-P30821).
\bibliography{ModelliCSL}

\end{document}